\documentclass[10pt]{article}

\usepackage{fullpage}
\usepackage{setspace}
\usepackage{parskip}
\usepackage{titlesec}
\usepackage{placeins}
\usepackage{xcolor}
\usepackage{float}
\usepackage{lineno}
\usepackage[toc,page]{appendix}
\usepackage{physics}
\usepackage[numbers,sort&compress]{natbib}

\usepackage{atbegshi}
\AtBeginDocument{\AtBeginShipoutNext{\AtBeginShipoutDiscard}}

\PassOptionsToPackage{hyphens}{url}
\usepackage[colorlinks = true,
            linkcolor = blue,
            urlcolor  = blue,
            citecolor = blue,
            anchorcolor = blue]{hyperref}
\usepackage{etoolbox}
\makeatletter
\patchcmd\@combinedblfloats{\box\@outputbox}{\unvbox\@outputbox}{}{%
}%
\makeatother

\renewenvironment{abstract}
  {{\bfseries\noindent{\large\abstractname}\par\nobreak}}

\titlespacing{\section}{0pt}{*3}{*1}
\titlespacing{\subsection}{0pt}{*2}{*0.5}
\titlespacing{\subsubsection}{0pt}{*1.5}{0pt}

\usepackage{authblk}

\usepackage{graphicx}
\usepackage[space]{grffile}
\usepackage{latexsym}
\usepackage{textcomp}
\usepackage{longtable}
\usepackage{tabulary}
\usepackage{booktabs,array,multirow}
\usepackage{amsfonts,amsmath,amssymb}
\providecommand\citet{\cite}
\providecommand\citep{\cite}

\newif\iflatexml\latexmlfalse
\DeclareGraphicsExtensions{.pdf,.PDF,.png,.PNG,.jpg,.JPG,.jpeg,.JPEG}

\usepackage[utf8]{inputenc}
\usepackage[english]{babel}

\begin{document}
\newcommand{\Sq}[1]{ \frac{ #1 }{ \sqrt{2} } }
\title{Ultrafast Molecular Frame Quantum Tomography}

\author[1]{Luna Morrigan}
\affil[1]{Department of Chemistry and Physics, University of Mary Washington, Fredericksburg, Virginia 22401, USA }

\author[2]{Simon P. Neville}
\affil[2]{National Research Council Canada, 100 Sussex Drive, Ottawa, ON, K1A 0R6, Canada\\}

\author[1]{Margaret Gregory}
\thanks{Current Address: Department of Earth, Atmospheric, and Planetary Sciences, Massachusetts Institute of Technology, 77 Massachusetts Avenue, Cambridge, MA 02139, USA.}

\author[3,4]{Andrey E. Boguslavskiy}
\affil[3]{Department of Physics, University of Ottawa, 150 Louis Pasteur, Ottawa, ON, K1N 6N5, Canada\\}%
\affil[4]{Department of Chemistry, University of Ottawa, Ottawa, ON, K1N 6N5, Canada\\}

\author[3,5]{Ruaridh Forbes}
\affil[5]{Linac Coherent Light Source, SLAC National Accelerator Laboratory, Menlo Park, CA 94025, USA}

\author[2.6]{Iain Wilkinson}
\affil[6]{Institute for Electronic Structure Dynamics, Helmholtz-Zentrum
f\"{u}r Materialien und Energie Berlin, Hahn-Meitner-Platz 1, 14109 Berlin, Germany}

\author[2]{Rune Lausten}

\author[2,3,4,7]{Albert Stolow}
\affil[7]{NRC-uOttawa Joint Centre for Extreme and Quantum Photonics (JCEP), Ottawa, ON, K1A 0R6, Canada.\\}

\author[2,4]{Michael S. Schuurman}

\author[2]{Paul Hockett}

\author[1]{Varun Makhija}
 \thanks{vmakhija@umw.edu}

\vspace{-1em}

  \date{\today}

\begingroup
\let\center\flushleft
\let\endcenter\endflushleft
\maketitle
\endgroup

\begin{abstract}
We develop and experimentally demonstrate a methodology for a full molecular frame quantum tomography (MFQT) of dynamical polyatomic systems. We exemplify this approach through the complete characterization of an electronically non-adiabatic wavepacket in ammonia (NH$_3$). The method exploits both energy and time-domain spectroscopic data, and yields the lab frame density matrix (LFDM) for the system, the elements of which are populations and coherences. The LFDM fully characterizes electronic and nuclear dynamics in the molecular frame, yielding the time- and orientation-angle dependent expectation values of any relevant operator. For example, the time-dependent molecular frame electronic probability density may be constructed, yielding information on electronic dynamics in the molecular frame. In NH$_3$, we observe that electronic coherences are induced by nuclear dynamics which non-adiabatically drive electronic motions (charge migration) in the molecular frame. Here, the nuclear dynamics are rotational and it is non-adiabatic Coriolis coupling which drives the coherences. Interestingly, the nuclear-driven electronic coherence is preserved over longer time scales. In general, MFQT can help quantify entanglement between electronic and nuclear degrees of freedom, and provide new routes to the study of ultrafast molecular dynamics, charge migration, quantum information processing, and optimal control schemes.
\end{abstract}

\section{Introduction.}
Molecular quantum electronic dynamics~\cite{kouppel1984,tannor1985,yarkony1996, domcke1997,zewail2000,domcke2004} govern important natural processes, including photosynthesis~\cite{engel2007}, vision~\cite{schoenlein1991}, photochemistry~\cite{calegari2014,smith2018} and solar energy conversion~\cite{gessner2019}. Attosecond science probes population dynamics and coherences between electronic states~\cite{li2022, chang2020, zinchenko2021, ruberti2021, huppert2016, calegari2014, dixit2012,matselyukh2022}.  The former often involves conical intersections generated by strong non-adiabatic coupling between the electrons and nuclei~\cite{kouppel1984,domcke2004}, the fundamental mechanism of energy transfer between them~\cite{blanchet1999, underwood2008, smith2018}. In general, the nuclear motions inducing such dynamics involve either rotation or vibration. Nuclear-driven electronic coherences generated at conical intersections are sensitive probes of their local topography~\cite{kowalewski2015,arnold2018,neville2022}. Electronic coherences may play an important role in fundamental light-induced processes~\cite{remacle1998, engel2007,li2022, kobayashi2019, kaufman2020, blavier2021}, thus measuring and controlling such coherences is of broad interest~\cite{ranitovic2014,suominen2014, santra2014, despre2018,sandor2018, simmermacher2019,hermann2020,carrascosa2021,folorunso2021,dey2022,plunkett2022,matselyukh2022}. In general, measurement and control of electronic populations and coherences requires experimental determination of the time-dependent-electronic density matrix - a quantum tomography~\cite{fano1957,dariano2003}. The latter underlies aspects of the foundations of quantum mechanics \cite{vrakking2021,koll2022,blavier2022} and molecular quantum information processing~\cite{akoury2007}. While probability distributions (static and dynamical) have been measured ~\cite{bisgaard2009time,schmidt2012,underwood2015,glownia2016, dixit2017, xiong2022}, quantum tomography was only demonstrated in restricted cases: a ground state rotational wavepacket, a stationary vibrational state, and a dissociative vibrational state~\cite{dunn1995,skovsen2003,mouritzen2006,hasegawa2008,zhang2021}. Recently, we proposed a systematic method for determination of the time-evolving electronic Lab Frame Density Matrix (LFDM) from experimental data~\cite{gregory2022}. We present here the first time-resolved Molecular Frame Quantum Tomography (MFQT).\\
\section{Molecular Frame Quantum Tomography in NH$_3$}
 In this proof-of-concept demonstration, we resonantly excited a pair of electronic states in NH$_3$, non-adiabatically coupled by molecular frame (MF) rotation~\cite{allen1991, makhija2020}. MFQT was achieved by combining data from ultrafast time-resolved measurements~\cite{makhija2020} with that of high-resolution spectroscopy~\cite{hockett2009}. The resulting density matrix yields the time-resolved electronic probability distribution in the MF, as a function of lab frame orientation angles. We show that nuclear-driven charge distributions evolves differently at different MF orientations. Importantly, the observed aperiodic charge migration is direct evidence of an angle-dependent non-adiabatic coupling, the angular analog to vibrational-coordinate-dependent non-adiabatic coupling. In our example, the electronic coherence persists over the entire 5~ps window of the time-resolved experiment. Long-lived electronic coherences are rare~\cite{dey2022, despre2018}, offering new opportunities for quantum control of molecular electronic dynamics~\cite{dalessandro2021}, the study of electronic-nuclear entanglement ~\cite{vrakking2021,koll2022,blavier2021}, and the development of quantum information processing protocols in isolated molecules~\cite{akoury2007}, for which quantum tomography is a necessary prerequisite.\\
\begin{figure}
    \centering
    \includegraphics[width = 1\columnwidth,trim=8 8 6 8,clip]{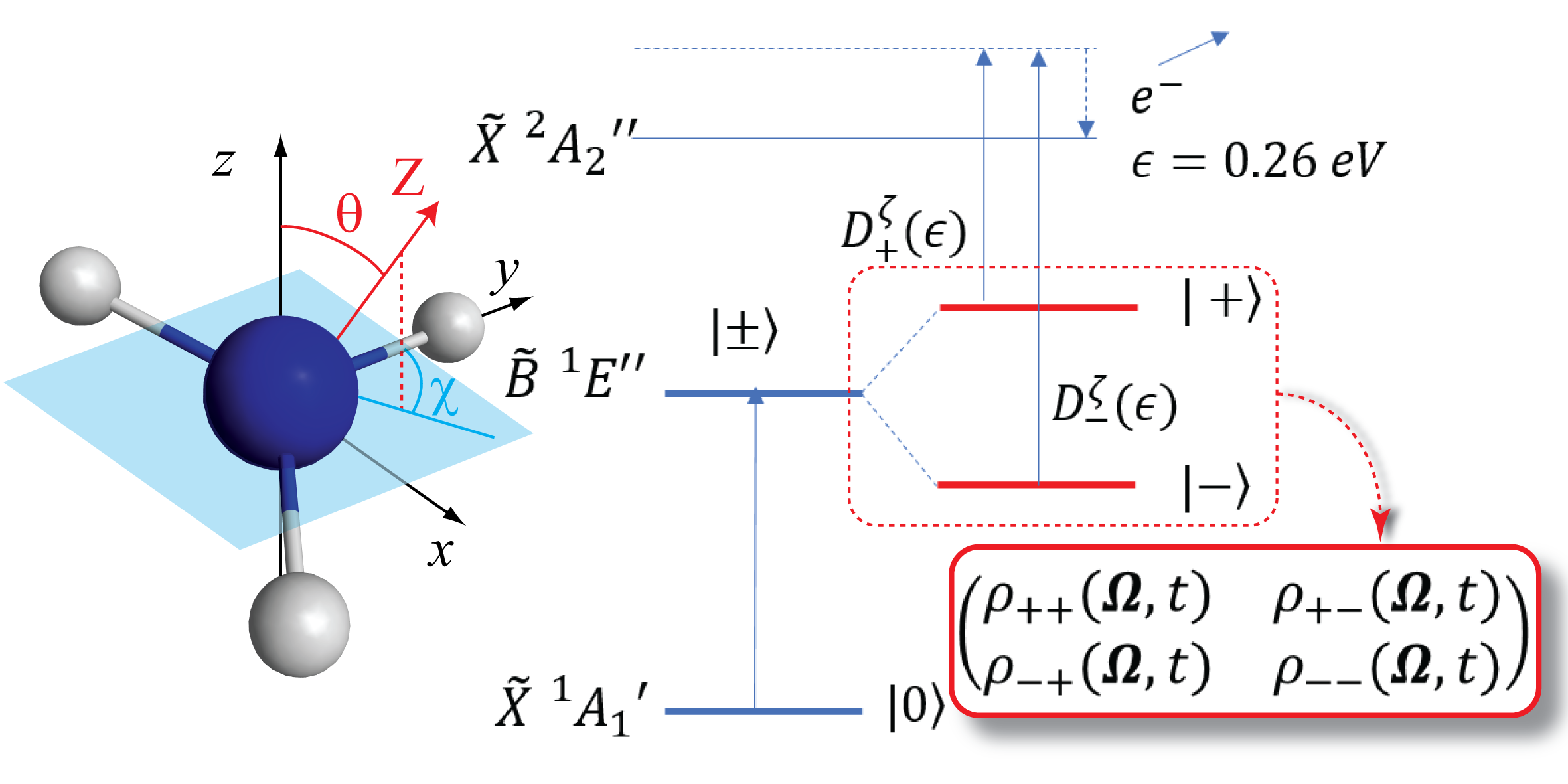}
    \caption{Electronic coherences non-adiabatically driven by nuclear motion. NH$_3$ is excited to the $\tilde{B}^1E''$ state, a pair of near degenerate electronic states $\ket{\pm}$, the LF $Z$-axis being the laser polarization direction. This excitation results in the time-dependent LF Density Matrix $\rho_{nn'}(\mathbf{\Omega},t)$, with $n\rightarrow\pm$. The planar geometry of NH$_3$ is shown with MF symmetry axis $z$ at angle $\theta$ with respect to the LF $Z$-axis. In-plane rotation about $z$ is given by the angle $\chi$. Photoionization into the $\tilde{X}\ {}^2A^{''}_2$ ionic state by the dipole operators $D^{\zeta}_{n}(\epsilon)$ produces an electron with kinetic energy $\epsilon=0.26$~eV. Both states ionize to overlapping continuum channels, $\zeta$, permitting the detection of electronic coherences. As we show, rapid nuclear motion along the $\theta$ coordinate induces electronic coherences which drive charge migration in the MF. For details, see the text.}
    \label{fig1}
\end{figure} 
In Fig.~\ref{fig1}, we depict NH$_3$ resonantly excited from a thermal rotational state distribution in the ground $\tilde{X}\ {}^1A'_{1}$ electronic state, $\ket{0}$, to its doubly degenerate $\tilde{B}^1E''$ state, $\ket{\pm}$, with three quanta in the umbrella vibrational mode. We determine the $2 \times 2$ orientation-dependent LFDM in the $\ket{\pm}$ basis, where $\Lambda_z\ket{\pm}=\pm 1 \ket{\pm}$, and $L_z = \xi \Lambda_z$~\cite{watson1984}. $L_z$ is the component of the electronic orbital angular momentum along the MF z-axis, the 3-fold symmetry axis of NH$_3$, and $\xi = \pm\bra{\pm}L_z\ket{\pm}$. In general, matrix elements of the LFDM can be written as~\cite{gregory2022}
\begin{equation}
\rho_{nn'}(\mathbf{\Omega},t)=\sum_{KQS}A^{K}_{QS}(n,n';t) D^{K*}_{QS}(\mathbf{\Omega})
\label{eqn1}
\end{equation}
where $\mathbf{\Omega}=\{\phi,\theta,\chi\}$ are the MF Euler angles and $n, n'\rightarrow\pm$ indicate the coherently excited electronic states. Molecular Angular Distribution Moments (MADMs) $A^{K}_{QS}(n,n';t)$ specify the evolving excited state molecular dynamics~\cite{gregory2022} and the $D^K_{QS}(\mathbf{\Omega})$ are Wigner D-Matrix elements~\cite{zare1988}. MFQT is enabled by determination of all relevant MADMs from the experimental data. The MADMs are multipole moments of the LFDM which track the time varying anisotropy of each LF matrix element. Selection rules for linearly polarized light restrict us to MADMs with $K = 0, 2$, $Q=0$ and $S = 0,\pm 2$, $|S| \leq K$~\cite{gregory2022}. Furthermore, the symmetry of the $\tilde{B}^1E''$ state permits only three unique, non-zero MADMs: $A^0_{00}(+,+;t)=A^0_{00}(-,-;t)$, $A^2_{00}(+,+;t)=A^2_{00}(-,-;t)$ and $A^2_{02}(+,-;t)=A^2_{0-2}(-,+;t)$~\cite{watson1984, allen1991, makhija2020}. The MADMs $A^0_{00}(\pm,\pm;t)$ track the total population in each excited state, while the $A^2_{00}(\pm,\pm;t)$ track the alignment of the z-axis for the population in each state. The $A^2_{0 \pm 2}(\pm,\mp;t)$ track the orientation of the electronic coherence in the lab frame. Their critical relationship to the electronic dynamics is detailed below.\\
\section{Determining MADMs}
In both the time- and frequency-resolved experiments, the excited states were probed by single photon photoionization to the $\tilde{X}\ {}^2A_2''$ state of NH$_3^+$~\cite{makhija2020,hockett20071,hockett20072,hockett2009}. The time-resolved data lead us to the time-dependent LFDM and the temporal evolution of the MF charge density. Here, NH$_3$ was excited by a 160.9~nm, 77~fs pump pulse, and ionized by a time delayed 400~nm, 40~fs probe pulse~\cite{makhija2020}. The photoelectron angular distribution and kinetic energy spectrum were measured as a function of time delay. Spherical harmonic expansion of the signal as a function of electron ejection angles $\theta_e$ and $\phi_e$, $P(\theta_e, \phi_e,\epsilon,t)=\sum_{LM}\beta_{LM}(\epsilon,t)Y_{LM}(\theta_e,\phi_e)$, provides the time- and electron-kinetic-energy ($\epsilon$)-dependent anisotropy parameters $\beta_{LM}(\epsilon,t)$. With linearly polarized pump and probe pulses, each a one-photon process, the three non-zero anisotropy parameters are $\beta_{00}$, $\beta_{20}$ and $\beta_{40}$~\cite{suzuki2006,underwood2008,Reid2012}. These, in turn, are expressed in terms of the MADMs~\cite{underwood2008,hockett2018QMP1,hockett2018QMP2},
\begin{equation}
\beta_{LM}(\epsilon, t) = \sum_{KQS}\sum_{nn'}C^{LM}_{KQS}(n,n';\epsilon)A^K_{QS}(n,n';t).
\label{eqn2}
\end{equation}
 Since the pump also generates three, unique non-zero MADMs then, with known coefficients $C^{LM}_{KQS}(n,n';\epsilon)$, Eq.~\ref{eqn2} becomes a matrix equation with solution $\Vec{A}(t)=\hat{C}^{-1}\Vec{\beta}(t)$ at each time delay. NH$_3$ is well-studied spectroscopically~\cite{Ashfold1988, douglas1963, Pratt2002, Reiser1993, softley2001}: the coefficients $C^{LM}_{KQS}(n,n';\epsilon)$ comprising $\hat{C}$ were previously determined by high-resolution Resonant Enhanced Multiphoton Ionization (REMPI) spectroscopy~\cite{hockett20071,hockett20072,hockett2009}. The coefficients can be written as $C^{LM}_{KQS}(n,n';\epsilon)=\sum_{\zeta\zeta'}\Gamma_{KQS}^{\zeta\zeta'LM}d^{nn'}_{\zeta\zeta'}(\epsilon)$ with $d^{nn'}_{\zeta\zeta'}(\epsilon) = D^n_{\zeta}(\epsilon)D^{n'*}_{\zeta'}(\epsilon)$. The factors $\Gamma_{KQS}^{\zeta\zeta'LM}$ are analytical and their properties were previously discussed at length~\cite{gregory2021}. The $D^n_{\zeta}(\epsilon)$ are matrix elements of the dipole operator between the bound state labeled $n$ and a continuum channel $\zeta$, specifying the final state of the ion plus free electron with kinetic energy $\epsilon$. Some partial wave matrix elements were determined for several electron kinetic energies, constituting a `complete experiment'~\cite{reid1992,motoki2002,lebech2003,cherepkov2005,teramoto2007,hockett2018QMP1,hockett2018QMP2}. Here, we use the results for $\epsilon = 0.26$~eV, relevant to our time-resolved experiments. Eq.~\ref{eqn2} is valid for $D_{3h}$ dipole matrix elements, symmetry adapted the point group of NH$_3$ in its $\tilde{B}^1E''$ state.
The $\Vec{\beta}(t)$ from the time-resolved and $\hat{C}$ from the frequency-resolved experiment (with associated experimental uncertainties) determine the MADMs\footnote{See Appdendices for a discussion of transformation to the symmetry adapted basis, plots of the MADMs, and construction of the MF charge density; which includes Refs.~\cite{underwood2008,hockett2009,makhija2020,gregory2021,gregory2022,hockett2018QMP1,hockett2018QMP2,bunkerjensen,child,ford2014,ashfold1998,allen1991,Ashfold1988,zare1988}}. The normalization $\sum_n A^0_{00}(n,n;t) = 1/8\pi^2$, equivalent to $\Tr{\rho(t)}=1$~\cite{gregory2022}, was applied at the initial time point and we rescaled the $K>0$ MADMs such that the ratio $A^K_{0S}/A^0_{00}$ remains unchanged. The resulting MADMs track the time varying population and molecular orientation in each electronic state and, critically, the coherence between them. These construct the LFDM $\rho(\mathbf{\Omega},t)$ in Eq.~\ref{eqn1} for any MF orientation $\mathbf{\Omega}$.\\
\begin{figure}
    \centering
    \includegraphics[width = 1\columnwidth]{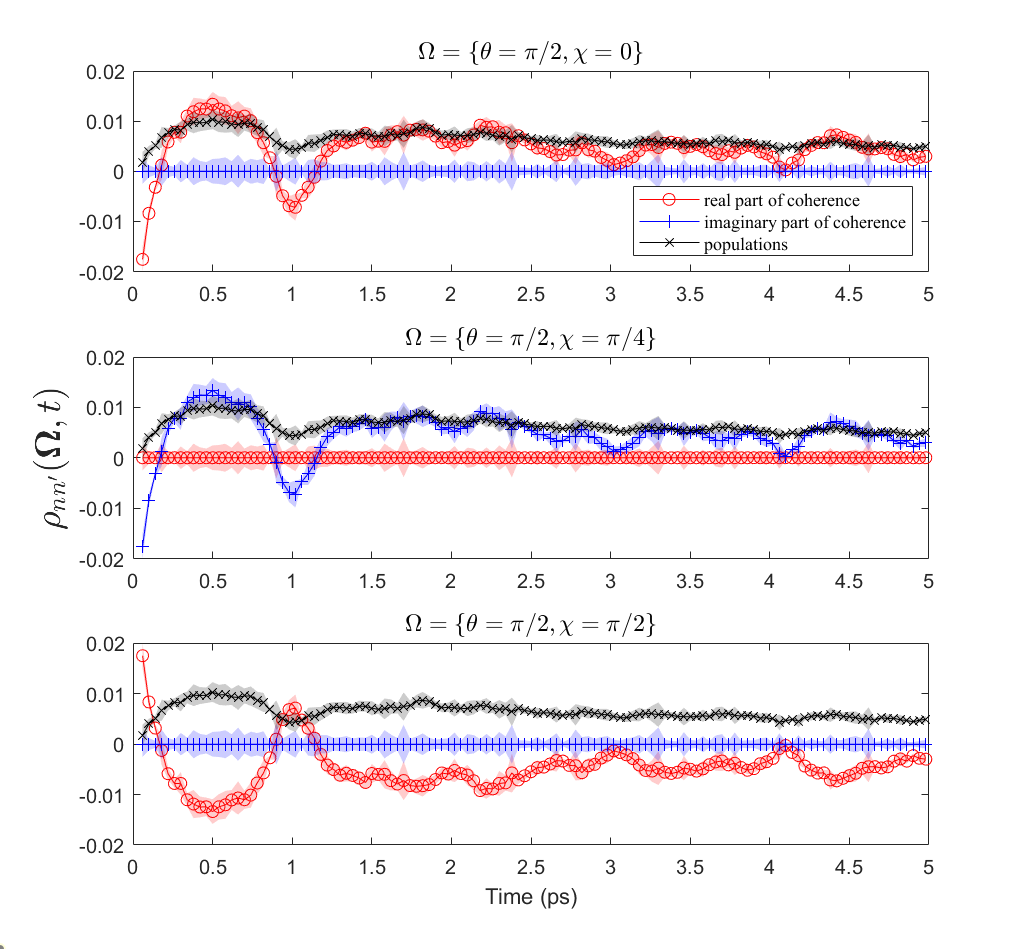}
    \caption{Experimentally determined elements of the time-resolved LFDM, $\rho_{nn'}(\mathbf{\Omega},t)$, for a molecule with $z$-axis perpendicular to the laser polarization (i.e, $\theta=\pi/2$), for different in-plane rotation angles (see Fig.~\ref{fig1}) $\chi=0$ (top), $\pi/4$ (middle) and $\pi/2$ (bottom). The electronic populations $\rho_{\pm\pm}(\{\pi/2,\chi\},t)$, black crosses, are independent of $\chi$; they initially increase then steadily decay, tracking the population of molecules oriented at $\theta=\pi/2$. In contrast, the electronic coherences $\rho_{+-}(\{\pi/2,\chi\},t)$ vary with $\chi$ and are the dominant contribution to the charge migration dynamics. They are real but counter-phased for $\chi = 0$ (top) and $\pi/2$ (bottom), indicating complementary electronic dynamics at these orientations. They are imaginary at $\chi = \pi/4$ (middle), revealing completely different electronic dynamics as a function of the nuclear coordinate $\chi$.}
    \label{fig3}
\end{figure}
\begin{figure}
	\includegraphics[width=0.9\columnwidth]{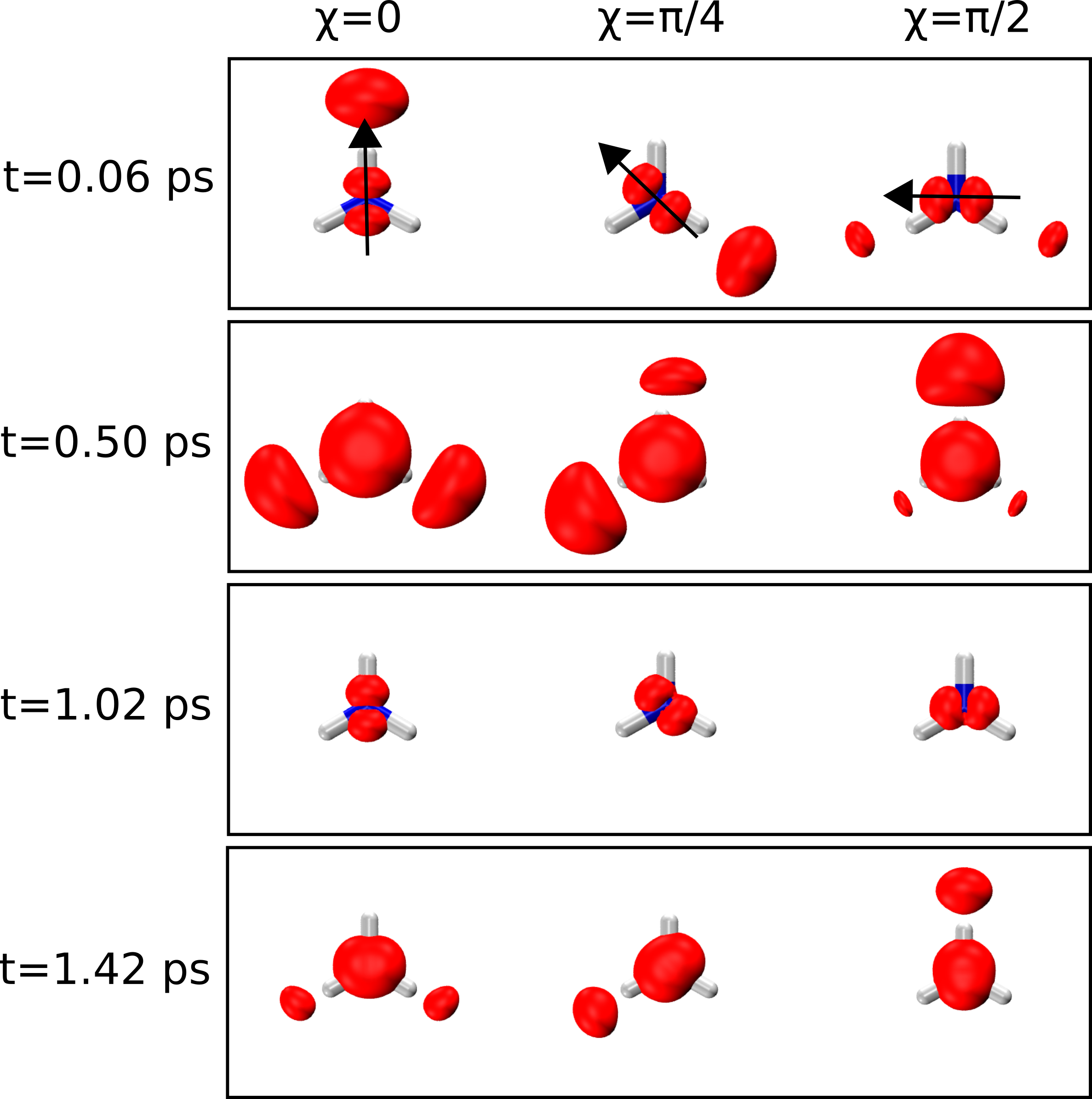}
	\caption{Nuclear-driven MF charge migration, extracted from the experimentally determined $\rho(\mathbf{\Omega},t)$. To illustrate, we show three columns depicting the time evolving attachment density $p_A(r_{1},\mathbf{\Omega},t)$ which tracks the variation of MF electron density, at three selected orientations; $\mathbf{\Omega} = \{\theta,\chi\}  = \{\pi/2,0\}$ (left),  $\{\pi/2,\pi/2\}$ (middle),  $\{\pi/2,\pi/2\}$ (right). The black arrow indicates the laser polarization direction $Z$ (see Fig.~\ref{fig1}). The electronic density evolves differently, and aperiodically, as a function of MF orientation, demonstrating nuclear coordinate-dependent non-adiabatic coupling between electronic states~\cite{patchkovskii2012,ruberti2021}, the angular analog of the well known vibrational-coordinate-dependent non-adiabatic coupling.}
	\label{fig4}
\end{figure} 
\section{Probing non-adiabatic dynamics}
We consider the electronic dynamics induced by nuclear motions. Selected elements of the extracted LFDM are plotted, at selected MF orientations, in Fig.~\ref{fig3}. We note that the diagonal elements tracking populations, $\rho_{++}(\mathbf{\Omega},t)=\rho_{--}(\mathbf{\Omega},t)$, are the same for all three orientations (black crosses), increasing in the first 0.5~ps and then slowly decaying. This indicates a higher probability for MF orientations with $\theta=\pi/2$ after 0.5~ps independent of $\chi$. The observed asymptotic behaviour is expected for a perpendicular pump transition and a symmetric top geometry~\cite{Ashfold1987,Ashfold1988,allen1991}. MF electronic dynamics at any orientation are dominated by the coherence $\rho_{+-}(\mathbf{\Omega},t)$. The real part of $\rho_{+-}(\mathbf{\Omega},t)$ is counter-phased for molecules oriented with $\chi = 0$ (top) and $\chi = \pi/2$ (bottom),  the imaginary part being zero. The electronic density at these orientations exhibit the complementary time evolution seen in the top and bottom rows. In contrast, at $\chi = \pi/4$ (middle) the real part of the coherence is zero, yielding entirely different electronic dynamics at this orientation.

From the orientation-dependent LFDM elements of Fig.~\ref{fig3}, we construct the MF one-electron reduced density,
\begin{equation}
p(r_{1},\mathbf{\Omega},t)=\sum_{nn'}\rho_{nn'}(\mathbf{\Omega},t)\int d{r}_{2} \cdots d{r}_{N} \psi_n^{*}(\Vec{r})\psi_{n'}(\Vec{r}),
\end{equation}
where $\Vec{r}=\{r_i|i=1,2,\dots,N\}$ is the set of position vectors of the electrons and $\psi_n(\Vec{r})$ is the wavefunction for a basis state $\ket{\pm}$. This yields a one-electron attachment density, $p_{A}(\Vec{r}_{1},\mathbf{\Omega},t)$, shown in Fig.~\ref{fig4}, depicting the orientation- and time-dependent accumulation of MF electron density relative to the static reference ground electronic state (for details see Appendices). To be consistent with common usage, we will refer hereafter to the observed MF evolution of the attachment density as `charge migration'~\cite{despre2018,sandor2018,simmermacher2019,hermann2020,carrascosa2021,folorunso2021,matselyukh2022}, but use this term to include both vectorial (directional) and tensorial (polarization) moments of the electronic dynamics. The left and right columns show the attachment density migrating along the $y$-axis for the orientations $\mathbf{\Omega}=\{\pi/2,0\}$ and $\mathbf{\Omega}=\{\pi/2,\pi/2\}$, but in opposing directions. Comparing these with the coherences at $\mathbf{\Omega}=\{\pi/2,0\}$ and $\mathbf{\Omega}=\{\pi/2,\pi/2\}$ of Fig.~\ref{fig3} reveals the correlation between the coherences and the MF electronic dynamics. At $\chi=0$, as the coherence first increases (between 0 and 0.5~ps), the density migrates downward, reversing as the coherence subsequently decreases. Interestingly, at $\chi = \pi/4$, the one-electron attachment density migrates around the $z$-axis. The radial extent of the plotted electronic density at all three orientations tracks the evolving population of perpendicularly oriented ($\theta=\pi/2$) molecules.
\begin{figure}
	\includegraphics[width=1\columnwidth]{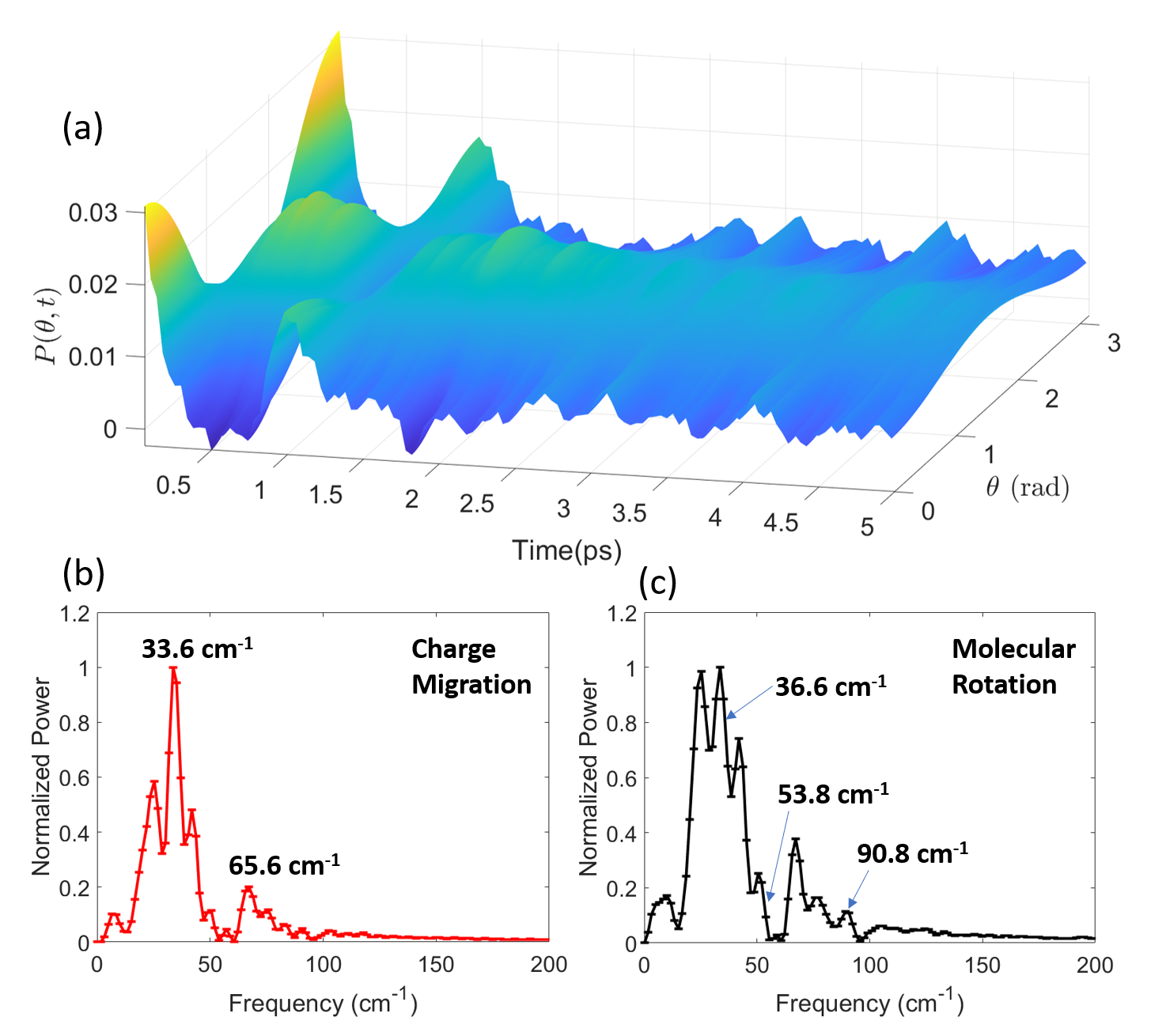}
	\caption{Experimentally determined nuclear-driven electronic coherences in NH$_3$. (a) The molecular Z-axis distribution $P(\theta,t)$, determined from the experimental LFDM, characterizes the excited state rotational dynamics. It can be seen that the nuclear coordinate $\theta$ varies rapidly at first but slows down at later times; (b) Electronic coherences and charge migration. Power spectrum of the real part of the $\rho_{+-}(\{\pi/2,0\},t)$. The dominant frequency $33.6\pm0.2$~cm$^{-1}$ (side bands at $25.2\pm0.2$~cm$^{-1}$ and $42.0\pm0.2$~cm$^{-1}$). The overtone appears at 65.6$\pm$0.2~cm$^{-1}$. These determine the timescales of the nuclear-induced charge migration in the MF; (c) Rotational dynamics. Power spectrum of $P(0,t)$ with the locations of expected symmetric top rotational frequencies based on the de-perturbed spectrum~\cite{ashfold1998}. It can be seen that the quantum beats cannot be classified as either rotational or electronic, rendering the motions inseparable. See text for additional details.}
	\label{fig5}
\end{figure} 
A nuclear coordinate-dependent aperiodic migration of electronic density in the MF is direct evidence of non-adiabatic dynamics~\cite{patchkovskii2012, ruberti2021}. Since we excite a single vibrational state, the nuclear dynamics of relevance here are rotational. Using the LFDM, we construct the time-varying molecular axis distribution, $P(\theta,t)=\sum_{n}\rho_{nn}(\theta,t)$, plotted in Fig.~\ref{fig5}(a), revealing the rotational dynamics, which are independent of $\chi$ as expected for a symmetric top~\cite{Ashfold1987}. In the first 1.5~ps, the most probable MF orientation oscillates rapidly between $\theta = 0$ and $\theta = \pi/2$. The electronic dynamics in Fig.~\ref{fig4} appear in this same time interval: the electronic coherence in Fig.~\ref{fig3} simultaneously exhibits large variations. Rapid nuclear motion, rotation of the MF $z$-axis, at early times drives the electronic coherence and, therefore, the charge migration in the MF. The power spectrum of the electronic coherence is shown in Fig.~\ref{fig5}(b), with peaks at $33.6\pm0.2$~cm$^{-1}$ and its overtone, providing the timescale for MF charge migration: $0.99\pm0.3$~ps. Later small fluctuations in $P(\theta,t)$ explain the persistent coherence. Beyond 1.5~ps, the most probable orientation remains relatively stable around $\theta = \pi/2$, with the time-averaged axis distribution peaking at $\theta = \pi/2$, as expected for a perpendicular transition~\cite{Ashfold1987,Ashfold1988,ashfold1998}. Later frames of the MF electron density (see Appendices) show that the density remains localized, with only small fluctuations. Slower fluctuations of molecular alignment (slower MF $\theta$ rotation) at later times thus stabilizes the electronic coherence. Furthermore, due to Coriolis coupling, frequency components of the electronic coherence also appear in the power spectrum of $P(0,t)$, shown in Fig.~\ref{fig5}(c). $P(0,t)$ also exhibits contributions, shown in Fig.~\ref{fig5}(c), near expected locations of rotational quantum beats, determined assuming a symmetric top Hamiltonian~\cite{makhija2020,ashfold1998}. Nevertheless, non-adiabatic coupling renders the electronic and rotational degrees of freedom non-separable. Therefore, all observed LF frequencies must be classified as quantum beats between ro-electronic molecular eigenstates. MFQT allows assignment of such beats by revealing which specific set of dynamical effects they contribute to in the LF and MF. 

MFQT reveals the dynamics underlying non-adiabatic nuclear-driven electronic coherences. In this proof-of-concept example, the nuclear dynamics are rotational, with Coriolis coupling driving the non-adiabaticity~\cite{ashfold1998}. Specifically, we note that: (i) the rotational and electronic dynamics, separable in the cation-plus-free-electron final state~\cite{makhija2020}, are non-separable in the excited state; (ii) rapid rotation of the MF $z$-axis ($\theta$) at early times drives a dynamic MF charge migration with a $\sim1$~ps period; (iii) subsequent small fluctuations of the MF $z$-axis preserve the electronic coherence over a long time. We emphasize that all this information is extracted from the experiment, without resorting to \emph{ab intio} dynamical simulations~\cite{nakamura1991,mayer1996}.

\section{Conclusions}\textemdash
We conclude by considering limitations and future applications of MFQT to complex molecules, charge migration and quantum control, and foundational quantum mechanics in molecules. A clear limitation is that, in determining the LFDM, $\rho_{nn'}(\mathbf{\Omega},t)\equiv\bra{\mathbf{\Omega}n}\rho\ket{n'\mathbf{\Omega}}$, we do not determine matrix elements of the density operator off-diagonal in the orientation angles, $\mathbf{\Omega}$. While this fully characterizes the electronic and vibrational dynamics in the MF, LF information is missing. We can construct the molecular axis distribution, but not observables sensitive to quantum coherences between different orientations, $\bra{\mathbf{\Omega}n}\rho \ket{n'\mathbf{\Omega}'}$, in the LF. Such observables are difficult to conceive since measurements relying on MF multipole interactions (like photoionization) are diagonal in $\ket{\mathbf{\Omega}}$ by definition. The von Neumann Entropy, $S=-\Tr{\rho\log{\rho}}$, is one quantity containing these coherences and thus cannot be constructed here.\\
There remain important avenues of investigation. The entanglement entropy of the vibronic subsystem, $S_{vib}(t) = -\Tr{\tilde{\rho}(t)\log{\tilde{\rho}(t)}}$, where $\tilde{\rho}_{nn'}(t)=8\pi^2A^{0}_{00}(n,n';t)$ is the reduced vibronic density matrix, can be constructed. The time-varying electron entropy, $S_{el}(t)$ in the NH$_3$ $\tilde{B}^1E''$ state may provide a quantitative measure of the electronic-rotational entanglement~\cite{amico2008,blum2012, tichy2011,goold2016}: its time-dependence may illuminate the role of entanglement in molecular electronic dynamics~\cite{vrakking2021,blavier2022,koll2022,rivera2022}. Investigating entanglement with an initially thermalized subsystem, as in this example, is an interesting prospect from the perspective of quantum thermodynamics~\cite{amico2008, goold2016, halpern2020}. 
Opportunities for Optimal quantum control of MF dynamics via $\rho(\mathbf{\Omega},t)$ also emerge~\cite{dalessandro2021}. For instance, in NH$_3$, the $\ket{\pm}$ states may be controlled by the non-resonant Dynamic Stark Effect\cite{sussman2006, ott2013, lewenstein2021}. Manipulating the LFDM in such a manner would control the time dependence of the electron density, a feature directly relevant to the burgeoning field of ultrafast molecular chirality~\cite{beaulieu2018, smirnova2009}. MFQT would allow similar experimental manipulation of charge migration in molecules, since the MF charge dynamics are directly accessible experimentally.\\
Photoinization-based MFQT requires as input complete REMPI experiments achieved only for a handful of molecules~\cite{reid1992,hockett20071,hockett20072,hockett2009}. Emerging attosecond techniques may be applicable: rotational wavepacket studies \cite{marceau2017} or angle-resolved RABBIT \cite{Laurent2012,hockett2017AngleresolvedRABBITTTheory} may provide sufficient information for \emph{in situ} complete photoionization experiments from an electronic molecular wavepacket. In general, when many electronic and/or vibrational states are excited, the matrix inversion problem in Eq.~\ref{eqn2} becomes ill-posed. Sophisticated mathematical methods were developed to deal with such situations, if physical constraints can be provided~\cite{ford2014,li2018}. Although the complete photoionization experiment problem itself can be similarly ill-posed, only products of the dipole matrix elements are needed to determine $\hat{C}$, circumventing the more complex problem of determining individual dipole matrix elements~\cite{gregory2021}. High quality \emph{ab initio} dipole matrix elements \cite{Lucchese1986,dowek2012PhotoionizationDynamicsPhotoemission,brown2020RMTRmatrixTimedependence,dowek2022TrendsAngleresolvedMolecular} may provide another suitable methodology. Finally, other angle-resolved scattering probes sensitive to the MADMs also apply~\cite{jonas2003,natan2021,hegazy2022}, provided the link between experiment and the MADMs is rigorously determined. We anticipate that this work will inspire a number of interesting directions in the study of quantum dynamics, charge migration, coherences and entanglement in isolated molecules.

\textbf{\emph{Acknowledgments}}\textemdash We thank Khabat Heshami and Serguei Patchkovskii for useful insights. VM thanks Vinod Kumarappan and Carlos Trallero for interesting discussions. RF acknowledges the US Department of Energy, Office of Science, Office of Basic Energy Sciences contract no. DE-AC02-76SF00515. AS and MSS thank the NRC-CSTIP Quantum Sensors grant \#QSP-075-1, and the NSERC Discovery Grants program for financial support. AS thanks the NRC-uOttawa Joint Centre for Extreme Photonics (JCEP) for financial support. 

\begin{appendices}
\section{Transformation from the partial wave to the symmetry adapted basis} 

\begin{figure}[hbt!] 
    \centering
    \includegraphics[width = .8\columnwidth,trim=8 8 6 8,clip]{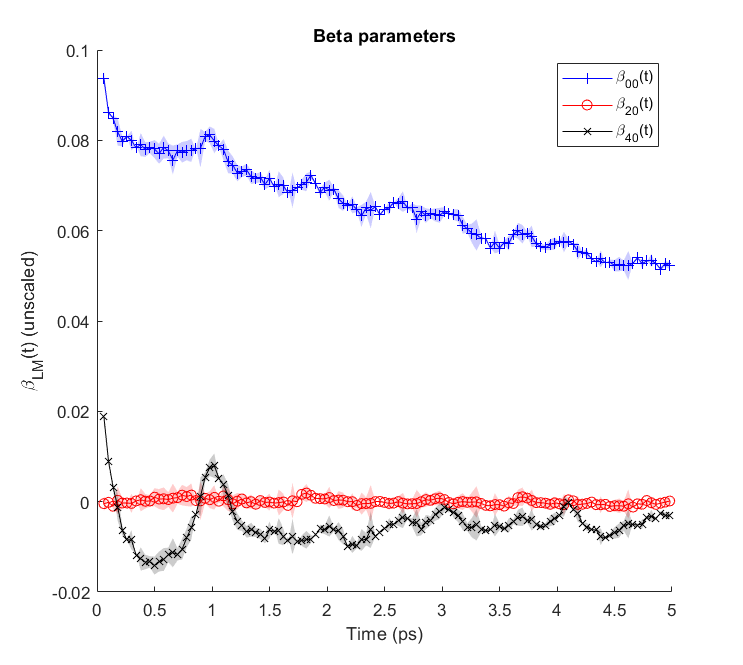}
    \caption{Measured $\beta_{LM}(t)$ parameters at a photoelectron kintic energy of $\epsilon=0.26$~eV~\cite{makhija2020}, with the shaded regions representing the statistical uncertainty.}
    \label{fig1_a}
\end{figure} 

\begin{table}[hbt!] 
\centering
\begin{tabular}{||c | c | c | c | c||}
\hline
$l$ & $\lambda$ & $q$ & $n$  & $D^n_{l\lambda q}$ \\
\hline \hline
0 &    0 & $\pm1$ & $\mp$ & 0.357 $\pm$ .0012 \\
\hline
1 & $\pm1$ & $\mp1$ & $\pm$ & 0.361 $\pm$ .0080\\
\hline
2 &    0  & $\pm1$ & $\mp$  &  0.1174 + .0706i $\pm$ .0219  \\
  & $\pm1$  & $\mp1$ & $\mp$  &  .3720 + .1067i $\pm$ .0204 \\  
  & 2  & 0,$\pm1$ &  $+$  &  -.5152 + .3095i $\pm$ .0212 \\
  & -2  & 0,$\pm1$ &  $-$  &  -.5152 + .3095i $\pm$ .0212 \\
  \hline
3 & $\pm1$ & $\mp1$ & $\pm$ & -.0799 + .0260i $\pm$ .0045\\

  & 2  & 0,$\pm1$ &  $+$  & -.1274 + .0649i $\pm$ .0027 \\
  & -2  & 0,$\pm1$ &  $-$  & -.1274 + .0649i $\pm$ .0027 \\
  
  & 3  & 0,$\pm1$ &  $+$  &  -.2513 + .1280i $\pm$ .0050 \\
  & -3  & 0,$\pm1$ &  $-$  &  -.2513 + .1280i $\pm$ .0050 \\
  \hline
4  &    0  & $\pm1$ & $\mp$  &  -.0060 + .1709i $\pm$ .0278 \\

  & $\pm1$ & $\mp1$ & $\pm$ &   .1210 + .2481i $\pm$ .1111 \\
  \hline
\end{tabular} \\
\caption{Photoionization dipole matrix elements in the partial wave basis from~\cite{hockett2009} at $\epsilon=0.26$~eV. The assignment of $n$ and $q$ is discussed in the text. }
\label{tab:Hockett2}
\end{table} 

The $\beta_{LM}(\epsilon,t)$ parameters measured in the time resolved experiment discussed in the main manuscript, originally published in Ref.~\cite{makhija2020}, are shown in Fig.~\ref{fig1_a}. The $C_{KQS}^{LM}(n,n';\epsilon)$ coefficients constituting the $\beta_{LM}(\epsilon,t)$ parameters, as in Eq.~2 of the main manuscript, fully describe the photoionization step used to probe the electronic dynamics. They are given by~\cite{gregory2022,gregory2021,makhija2020,underwood2008},
\begin{equation}
C^{LM}_{KQS}(n,n';\epsilon) = \sum_{\zeta\zeta'} D^{n}_\zeta(\epsilon) D^{n'*}_{\zeta'}(\epsilon) \Gamma^{\zeta\zeta'LM}_{K0S},
\label{CLM}
\end{equation}
where the $D^n_\zeta(\epsilon)$ are photoionization dipole matrix elements discussed in the manuscript and the $\Gamma^{\zeta\zeta'LM}_{K0S}$ are previously studied analytical factors~\cite{gregory2021,hockett2018QMP1,hockett2018QMP2}. The dipole matrix elements for photoionization of the $B^1E''$ state of NH$_3$  in the partial wave basis, $\zeta \rightarrow l, \lambda, q$, were determined by Resonance Enhanced Multiphoton Ionization~\cite{hockett2009}, and are provided in table~\ref{tab:Hockett2}, for a photoelectron kinetic energy $\epsilon = 0.26$~eV as measured in the time resolved experiment. We suppress the $\epsilon$ argument from here on. In this basis, the wavefunction of the ejected electron is expanded in spherical harmonics, $Y_{l\lambda}(\theta_e,\phi_e)$, where $\theta_e$ and $\phi_e$ are the spherical polar angles of the ejected electron. Furthermore, $q=0, \pm1$ labels the vector components of the dipole in the spherical basis and $n\rightarrow \pm$ labels the near-degenerate components of the $B^1E''$ state. These labels are determined by selection rules explained below. To use Eq.~\ref{CLM} the $D^n_\zeta$ must be transformed to a symmetry adapted basis $X^{\Gamma \mu}_{hl}$~\cite{underwood2008};
\begin{equation}
X^{\Gamma \mu}_{hl}(\theta_e,\phi_e) = \sum_{\lambda}b^{\Gamma \mu}_{h l \lambda}Y_{l\lambda}(\theta_e,\phi_e).
\end{equation}
The transformation coefficients $b^{\Gamma \mu}_{h l \lambda}$ may be determined using the projection operator for the representation $\Gamma$ of the $D_{3h}$ point group~\cite{bunkerjensen}. Expansions for several $l$'s are provided in table~\ref{tab:D1}. From these expansions we can assign the label $\mu\rightarrow |\lambda|$, and the `degeneracy index'~\cite{underwood2008} $h\rightarrow \pm$, labeling the sign of $\lambda$, which is only needed for the degenerate $E'$ and $E''$ representations. The electronic configuration of the $B^1E''$ state can also be written in this basis, with the highest occupied orbital generating the doubly degenerate $E'$ representation of $D_{3h}$~\cite{child}. As evident from Table~\ref{tab:D1} this representation preserves the sign of projection of $l$, and $l = 0$ is not allowed. We can therefore express the $\ket{\pm}$ states in a spherical harmonic expansion as follows,
\begin{equation}
\psi_{\pm}(r_e,\theta_e,\phi_e) = \sum_{l|m|}f_{l|m|}(r)Y_{l\pm|m|}(\theta_e,\phi_e).
\end{equation}
Using this expansion, and applying the Wigner-Ekart theorem~\cite{zare1988} to the matrix elements of the dipole operator $\mu^1_q$ in a spherical basis, $\bra{l'\lambda}\mu^1_q\ket{lm}$, provides the selection rules,
\begin{equation}
\begin{split}
l'=l\pm1\\ \lambda=m+q,
\end{split}
\end{equation}
linking $l, \lambda, q$ and $n$ as given in table~\ref{tab:Hockett2}. Note that these are still in the partial wave basis. Conversion to the symmetry adapted basis is easily accomplished by making the appropriate linear combinations of the $D^n_{l\lambda q}$ using the transformation coefficients $b^{\Gamma\mu}_{hl\lambda}$ given in table~\ref{tab:D1}. The allowed representations $\Gamma$ for the ejected electron are determined by the symmetry selection rule~\cite{underwood2008}
\begin{equation}
A_1' \subset \Gamma \otimes \Gamma_q \otimes \Gamma_i
\end{equation}
where $\Gamma_i = E'$ is the symmetry of the ionized orbital, $\Gamma_q$ the symmetry of the dipole component, and $A_1'$ the totally symmetric representation of $D_{3h}$. The allowed representations along with transformed values of $D_{\Gamma \mu h l q}^n$ are listed in table~\ref{tab:D1}.
\begin{table}[H]
\centering
\begin{tabular}{|| c | c | c | c | c | c | c||}
\hline
$\Gamma$ & $l$ & $\lambda$ & $q$ & $n$ & $X^{\Gamma \mu}_{h l}$ & $D^n_{\Gamma \mu h l q}$  \\
\hline\hline
$A_1'$& 0 & 0 &  $\pm1$ & $\mp$  & $Y_{00}$ & .357 $\pm$ .012 \\
      & 2 & 0 &  $\pm1$ & $\mp$    & $Y_{20}$ & .1174 + .0706i $\pm$ .0218  \\
      \hline
      & 3 & [3,-3] &  $\pm1$ &  + & $\Sq{1}Y_{33} - \Sq{1}Y_{3-3}$ & -.1777 +.0905i $\pm$ .0036 \\
      & 3 & [3,-3] &  $\pm1$ &  - & $\Sq{1}Y_{33} - \Sq{1}Y_{3-3}$ & .1777 -.0905i $\pm$ .0036 \\
      \hline
      & 4 & 0 &  $\pm1$ &  $\mp$ & $Y_{40}$ & -.0060 + .1709i $\pm$ .0278 \\     
\hline\hline
$A_2'$& 3 & [3,-3] &  $\pm1$ &  + & $\Sq{1}Y_{33} + \Sq{1}Y_{3-3}$ & -.1777 +.0905i $\pm$ .0036 \\
      & 3 & [3,-3] &  $\pm1$ &  - & $\Sq{1}Y_{33} + \Sq{1}Y_{3-3}$ & -.1777 +.0905i $\pm$ .0036 \\      
\hline\hline
$E'$& 1 &  $\pm1$ & $\mp1$ &  $\pm$ & $Y_{1\pm1}$  & .361 $\pm$ .0080\\
    \hline
    & 2 &  2 &  $\pm1$ &  + & $Y_{22}$  & -.5152 + .3095i $\pm$ .0212 \\   
    & 2 & -2 &  $\pm1$ &  - & $Y_{2-2}$ & -.5152 + .3095i $\pm$ .0212 \\   
    \hline
    & 3 &  $\pm1$ & $\mp1$ &  $\pm$ & $Y_{3\pm1}$  & .0799 +.0260i $\pm$ .0045\\   
\hline\hline
$E''$& 2 &  $\pm1$ & 0 &  $\pm$ & $Y_{2\pm1}$  & .3720 +1067i $\pm$ .0204 \\    
     \hline
     & 3 &  $\pm2$ & 0 &  $\pm$ & $Y_{3\pm2}$  & -.1274 + .0649i $\pm$ .0027 \\
     \hline
     & 4 &  $\pm1$ & 0 &  $\pm$ & $Y_{4\pm1}$  & .1210 + .2481i $\pm$ .1111 \\     
\hline\hline
\end{tabular}
\caption{The photoionization dipole matrix elements $D^n_{\Gamma\mu hlq}$ in the symmetry adapted basis $X^{\Gamma\mu}_{hl}(\theta_e,\phi_e)$. Only representations for dipole allowed transitions are considered. See the text for further details.}
\label{tab:D1}
\end{table}  

\begin{figure}[hbt!]
    \centering
    \includegraphics[width = 1\columnwidth]{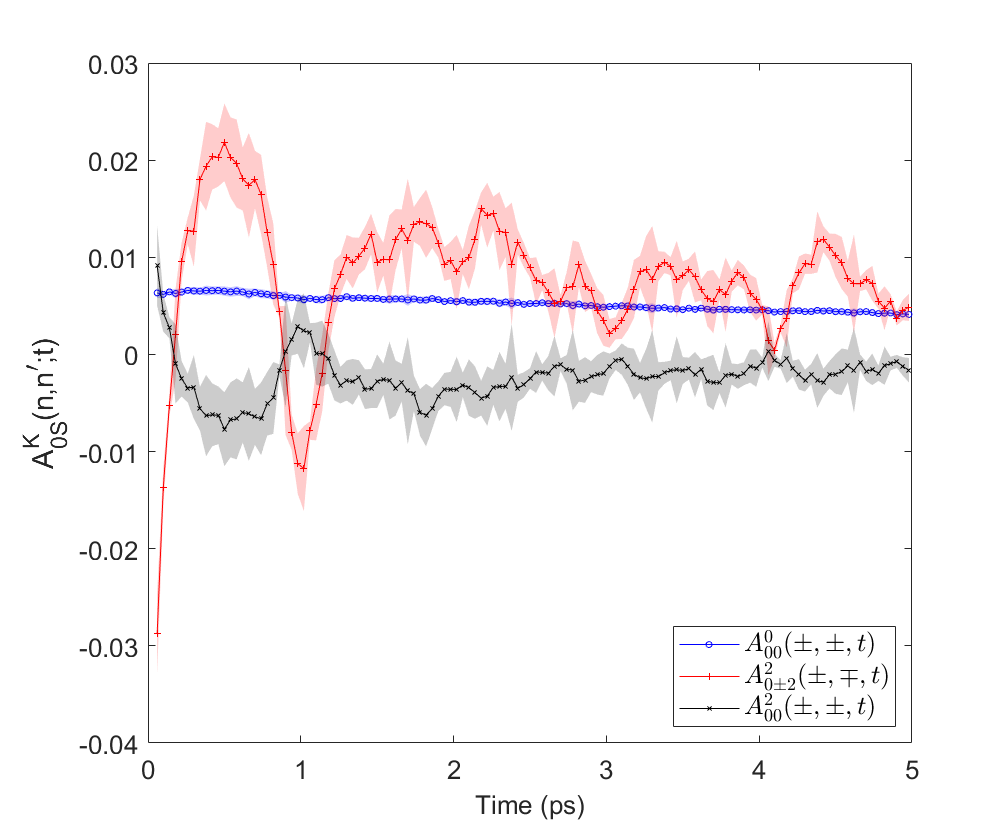}
    \caption{Time evolution of the experimentally determined MADMs for the excited superposition of the $\ket{\pm}$ states of NH$_3$ extracted from measured $\beta_{LM}(0.26\ \textrm{eV}, t)$ values~\cite{makhija2020} using a $\hat{C}$ matrix constructed from the dipole matrix elements at $\epsilon = 0.26$~eV ~\cite{hockett2009}. For details, see the text.} 
    \label{fig2_a}
\end{figure}

Finally, the $D^n_\zeta$ from table~\ref{tab:D1} are used in Eq.~\ref{CLM} to construct the $\hat{C}$ matrix,
\begin{equation}
\hat{C} = \begin{pmatrix} 0.9605\pm.0248 & -0.0935\pm.0088 & -0.0573\pm.0017\\ 0 & -0.0633\pm.0098 & -0.0190\pm.0023 \\ 0 & -0.0143\pm.0050 & -0.0526\pm.0034 \end{pmatrix}.
\end{equation} 
Here the rows correspond to the three measured time dependent traces $\beta_{00}(t)$, $\beta_{20}(t)$ and $\beta_{40}(t)$ in Fig.~\ref{fig1}; and the columns to the three unique MADMs $A^0_{00}(+,+;t) = A^0_{00}(-,-;t)$, $A^2_{00}(+,+;t) = A^2_{00}(-,-;t)$ and $A^2_{02}(+,-;t) = A^2_{0-2}(-,+;t)$. 
 The entries in the matrix  are linear combinations of the corresponding $C^{LM}_{KQS}(n,n';\epsilon)$ calculated using Eq~\ref{CLM} for $\epsilon = 0.26$~eV and the $D^n_\zeta$ from table~\ref{tab:D1}. We have set $C_{12} = 0$ since we subtract the mean of the measured $\beta_{20} (t)$ from each data point. This removes the DC component from $\beta_{20}(t)$, corresponding to the populations of the states, rendering $C_{12} = 0$, thus providing a simplified system of equations to the inversion algorithm. Finally, $\hat{C}$ is non-singular, such that $\hat{C}^{-1} = \textrm{adj}(\hat{C})/\textrm{det}(\hat{C})$ is defined, where $\textrm{adj}(\hat{C})$ and $\textrm{det}(\hat{C})$ are the adjoint and determinant of $\hat{C}$ respectively~\cite{ford2014}. Arranging the $\beta_{LM}(t)$ from Fig.~\ref{fig1} in a column vector $\vec{\beta}(t)$, then computing $\vec{A}(t) = \hat{C}^{-1}\vec{\beta(t)}$, provides the three unique MADMs in Fig.~\ref{fig2_a}. 

We observe that $A^{0}_{00}(+,+;t) = A^{0}_{00}(-,-;t)$ slowly decays, tracking the well-known loss of population to the $\tilde{A}\ {}^1A_2''$ state via internal conversion~\cite{Ashfold1988,allen1991}. $A^2_{00}(+,+;t)=A^2_{00}(-,-;t)$ tracks the molecular alignment in each state and is largely negative, indicating that the $z$-axis is preferentially aligned perpendicular to the polarization of the pump pulse ($\theta = \pi/2$), as expected for a perpendicular transition~\cite{ashfold1998,allen1991}. $A^{2}_{02}(+,-;t)=A^{2}_{0-2}(-,+;t)$ tracks the coherence, which, notably, is the largest contribution to the dynamics and remains so for the duration of the experiment. These MAMDs are used to construct the density matrix elements shown in Fig. 2 of the main manuscript.

\section{Calculation of molecular frame attachment densities}

\subsection{One-electron reduced densities}
The molecular frame (MF) one-electron reduced density,
$p(r_{1},\boldsymbol{\Omega},t)$, is obtained from the integration of
the $N$-electron MF electronic probability density

\begin{equation}
  P(\Vec{r},\boldsymbol{\Omega},t) = \sum_{nn'}
  \rho_{nn'}(\boldsymbol{\Omega},t) \psi_{n}^{*}(\Vec{r})
  \psi_{n'}(\Vec{r})
\end{equation}

\noindent
over all but one electronic degrees of freedom,

\begin{equation}
  p(r_{1}, \boldsymbol{\Omega}, t) = \sum_{nn'}
  \rho_{nn'}(\boldsymbol{\Omega},t) \int dr_{2} \cdots dr_{N}
  \psi_{n}^{*}(\Vec{r}) \psi_{n'}(\Vec{r}).
\end{equation}

\noindent
For the sake of clarity, in the following we shall drop the subscript
from $r_{1}$, with the understanding that `$r$' is to be equated with a
single (indistinguishable) electronic degree of freedom.

Introducing a single-electron (molecular orbital [MO]) basis $\{
\varphi_{i}(r) \}$ with associated elementary creation (annihilation)
operators $\hat{a}_{i}^{\dagger}$ ($\hat{a}_{i}$), the one-electron
reduced density may be written as

\begin{equation}
  p(r, \boldsymbol{\Omega}, t) = \sum_{nn'}
  \rho_{nn'}(\boldsymbol{\Omega}, t) \sum_{ij} D_{ij}^{(n,n')}
  \varphi_{i}^{*}(r) \varphi_{j}(r),
\end{equation}

\noindent
where the one-electron reduced (transition) density matrices
$\boldsymbol{D}^{(n,n)}$ ($\boldsymbol{D}^{(n,n')}, n \ne n'$) are
given by

\begin{equation}
  D_{ij}^{(n,n')} = \langle \psi_{n} | \hat{a}_{i}^{\dagger}
  \hat{a}_{j} | \psi_{n'} \rangle.
\end{equation}

\subsection{Visualization and attachment densities}
Let $\boldsymbol{D}(\boldsymbol{\Omega},t)$ denote the MO
representation of the MF one-electron reduced density:

\begin{equation}
  p(r, \boldsymbol{\Omega}, t) = \sum_{ij}
  D_{ij}(\boldsymbol{\Omega}, t) \varphi_{i}(r) \varphi_{j}(r),
\end{equation}

\noindent
with

\begin{equation}
  D_{ij}(\boldsymbol{\Omega}, t) = \sum_{nn'}
  \rho_{nn'}(\boldsymbol{\Omega}, t) D_{ij}^{(n,n')}.
\end{equation}

In general, the time- and angle dependencies of the MF one-electron
reduced density may be subtle and thus hard to discern visually. In
order to emphasize these differences, we may instead consider the MF
one-electron difference density $\Delta p(r, \boldsymbol{\Omega}, t)$,
obtained via the subtraction of a suitable constant reference density
$p_{0}(r)$, chosen here as that of the ground electronic state,

\begin{equation}
  \begin{aligned}
    \Delta p(r, \boldsymbol{\Omega}, t) &= p(r, \boldsymbol{\Omega},
    t) - p_{0}(r)\\
    &= \sum_{ij} \left[ D_{ij}(\boldsymbol{\Omega}, t) -
      D_{ij}^{(0,0)} \right] \varphi_{i}(r) \varphi_{j}(r) \\
    &= \sum_{ij} \Delta D_{ij}(\boldsymbol{\Omega}, t) \varphi_{i}(r)
    \varphi_{j}(r),
  \end{aligned}
\end{equation}

\noindent
where $\boldsymbol{D}^{(0,0)}$ denotes the ground state one-electron
reduced density matrix,

\begin{equation}
  D_{ij}^{(0,0)} = \langle \psi_{0} | \hat{a}_{i}^{\dagger}
  \hat{a}_{j} | \psi_{0} \rangle.
\end{equation}

The one-electron difference density $\Delta p(r, \boldsymbol{\Omega},
t)$ may be further decomposed into attachment and detachment
densities, $p_{A}(r, \boldsymbol{\Omega}, t)$ and $p_{D}(r,
\boldsymbol{\Omega}, t)$, respectively:

\begin{equation}
  \Delta p(r, \boldsymbol{\Omega}, t) = p_{A}(r, \boldsymbol{\Omega},
  t) + p_{D}(r, \boldsymbol{\Omega}, t),
\end{equation}

\begin{equation}
  p_{A}(r, \boldsymbol{\Omega}, t) = \sum_{i}
  \max(\lambda_{i}(\boldsymbol{\Omega},t),0) \left|
  \overline{\varphi}_{i}(r,\boldsymbol{\Omega},t) \right|^{2},
\end{equation}

\begin{equation}
  p_{D}(r, \boldsymbol{\Omega}, t) = \sum_{i}
  \min(\lambda_{i}(\boldsymbol{\Omega},t),0) \left|
  \overline{\varphi}_{i}(r,\boldsymbol{\Omega},t) \right|^{2}.
\end{equation}

\noindent
Here, the MF natural difference orbitals
$\overline{\varphi}_{i}(r,\boldsymbol{\Omega},t)$ are given by

\begin{equation}
  \overline{\boldsymbol{\varphi}}(r,\boldsymbol{\Omega},t) =
  \boldsymbol{U}(\boldsymbol{\Omega},t) \boldsymbol{\varphi}(r),
\end{equation}

\noindent
where $\boldsymbol{U}(\boldsymbol{\Omega},t)$ denotes the matrix of
eigenvectors of $\Delta \boldsymbol{D}(\boldsymbol{\Omega},t)$, with
corresponding eigenvalues $\lambda_{i}(\boldsymbol{\Omega},t)$.

The attachment (detachment) density corresponds to the regions of
accumulation (depletion) of electron density relative to the reference
state. In the specific system considered here, NH$_{3}$ excited to its
$\tilde{B}\ {}^1E''$ state, the detachment density is quasi-static,
exhibiting only small angular and time dependencies in terms of
magnitude. We thus chose to focus on the visualization of the
attachment density, which is found to encode the majority of the
dynamical information.

\subsection{Calculation details}
The one-electron reduced (transition) densities
$\boldsymbol{D}^{(n,n')}$ were computed at the combined density
functional theory and multi-reference configuration interaction
(DFT/MRCI) level of theory using the aug-cc-pVTZ basis set. The
required MOs were computed using the PySCF progam package, and the
DFT/MRCI calculations using the Generalized Reference Configuration
Interaction (GRaCI) code
\end{appendices}
         
\bibliographystyle{unsrt}
\bibliography{refs, NH3, MatEtheory}


\end{document}